\documentclass[twocolumn]{aastex62}

\usepackage{graphicx}	
\usepackage{amsmath}	
\usepackage{amssymb}	
\usepackage{soul}
\received{January 1, 2020}
\revised{January 7, 2020}
\accepted{\today}
\submitjournal{ApJ}

\shorttitle{Insights from snapshot spectroscopic radio observations of a weak Type I noise storm}
\shortauthors{Mondal \& Oberoi}

\begin{document}

\title{
Insights from snapshot spectroscopic radio observations of a weak Type I solar noise storm
}


\correspondingauthor{Surajit Mondal}
\email{surajit@ncra.tifr.res.in}

\author{Surajit Mondal}
\affil{National Centre for Radio Astrophysics, \\
Tata Institute of Fundamental Research, \\
Pune-411007, India}

\author{Divya Oberoi}
\affil{National Centre for Radio Astrophysics, \\
Tata Institute of Fundamental Research, \\
Pune-411007, India}



\begin{abstract}
We present a high fidelity snapshot spectroscopic radio imaging study of a weak type I solar noise storm which took place during an otherwise exceptionally quiet time.
Using high fidelity images from the Murchison Widefield Array, we track the observed morphology of the burst source for 70 minutes and identify multiple instances where its integrated flux density and area are strongly anti-correlated with each other. 
The type I radio emission is believed to arise due to electron beams energized during magnetic reconnection activity. 
The observed anti-correlation is interpreted as evidence for presence of MHD sausage wave modes in the magnetic loops and strands along which these electron beams are propagating.
Our observations suggest that the sites of these small scale reconnections are distributed along the magnetic flux tube.
We hypothesise that small scale reconnections produces electron beams which quickly get collisionally damped. 
Hence, the plasma emission produced by them span only a narrow bandwidth and the features seen even a few MHz apart must arise from independent electron beams. 
\end{abstract}


\section{Introduction}
\label{intro}

Type I noise storms are the most commonly observed solar nonthermal emission in the metric wavelengths. These emissions are generally associated with active regions and can persist on the solar disc for durations lasting up to several days \citep{elgaroy1977}. 
The solar noise storms are characterised by short duration ($\sim 0.1-10$s) intense narrowband ($\sim$ a few MHz) bursts superposed on a long duration wideband ($\gtrsim 100$ MHz) emission. The emission is strongly circularly polarised \citep[e.g.][etc.]{zlobec1971, ramesh2011,ramesh2013, mugundhan2018,mccauley2019}. These emissions are believed to be arising from intense plasma emission from nonthermal accelerated electrons trapped in the coronal loops \citep{melrose1980}. Imaging studies of Type I noise storms have revealed that these sources are quite compact and can have compact substructures as well \citep[e.g.][etc.]{mercier2015,mugundhan2018,mohan2019b}. Type I noise storms are generally associated with active regions \citep{sakurai1971} and often accompany energetic flares and CMEs \citep[e.g.][]{smith1962,kathiravan2007, dudik2014, mugundhan2018}. However, sensitive observations have also revealed their association with small scale magnetic enhancements and brightenings in the extreme ultraviolet (EUV) bands as well \citep[e.g.][]{iwai2012,li2017,mohan2019b}. 
There have also been suggestions that the persistent radio emission from Type I noise storms can be produced due to the persistent magnetic reconnections along quasi-separatrix layers of active regions \citep{zanna2011,mandrini2015}.


Active regions host a number of coronal loops, which in turn host a variety of MHD wave modes - like sausage mode and torsional modes \citep{aschwanden2005}.
These are believed to be excited by the stochastic and often persistent reconnection processes.
These waves are also known to influence the reconnection process itself \citep[e.g][]{carley2019, mohan2019,mohan2021}. 
Hence, it is natural to expect that these waves can influence the dynamics of a Type I noise storm associated with the active region.
The presence of these waves in Type I noise storms are generally inferred from observations of quasi-periodic pulsations (QPP) \citep[e.g.][]{sastry1969,sundaram2004}. For a detailed review regarding QPPs in general, we refer the reader to \citet{nakariakov2009}.

Detection of QPPs necessarily requires a dominant timescale to be present in the system for a significant time duration.
It is, however, possible to envisage systems where these MHD waves are produced only intermittently and for a small fraction of the observing duration, and hence do not give rise to an identifiable dominant timescale. 
Under such circumstances, despite their presence, it will not be possible to detect QPPs at high significance levels. 
If high-fidelity radio images are available, then even { in the absence of dominant timescale in the system }, it should be possible to infer the presence of MHD waves by looking for signatures which give rise to correlations between the different morphological parameters of type I source, without requiring one to establish the sustained presence of quasi-periodic { emission }features in the data. 
To the best of our knowledge, presence of waves was first inferred using such a technique of correlated evolution or morphological parameters by \citet{mohan2019}. 
They studied a type III solar radio burst, demonstrated the presence of a strong anti-correlation between area and integrated flux density of the burst source and interpreted it in terms of a MHD wave driven modulation  mechanism operating at the particle injection/acceleration site. 
More recently \citet{mohan2021} have analysed the observed correlated evolution in the morphological properties of the type I noise storm during an active region transient brightening (ARTB) associated with a microflare. They establish the presence of QPPs in the radio observables and interpret them as evidence for presence of sausage and torsional mode MHD waves during microflares.
Here, we present evidence for the presence of these waves during the quiescent phase of a type I noise storm, when its radio flux density is more than two orders of magnitude weaker than in the study by \citet{mohan2021}. 
Our work also demonstrates that even during a very quiet period, small scale reconnections continue to take place in the vicinity of active regions. 

{ Naturally, as this approach relies on  characterizing the morphology of a radio burst source to glean information about presence of QPPs, it is only feasible when such a source is available.
Hence, the presence of nonthermal electron beams, which set up instabilities as they propagate through the heliospheric plasma and give rise to { observable }radio emissions at the local plasma frequency (and/or its harmonic) yielding a suitable source, are an essential requirement for this approach. 
Here we study a type I noise storm, which owes its existence to the presence of nonthermal electron beams for the duration of the storm.
}

This paper is organised as follows -- Sec. 2 describes the observations and the data analysis. In Sec. 3, we describe the results, present their discussion in Sec. 4, which is followed by the conclusions in Sec. 5.

\section{Observations and Data Analysis}
\label{sec:obs}
These data have been described in detail by \citet{mondal2020}, referred to subsequently as M20. 
Here we only present {some key aspects of interest} about the observations and the data analysis.

\begin{figure}
    \centering
    \includegraphics[scale=0.5]{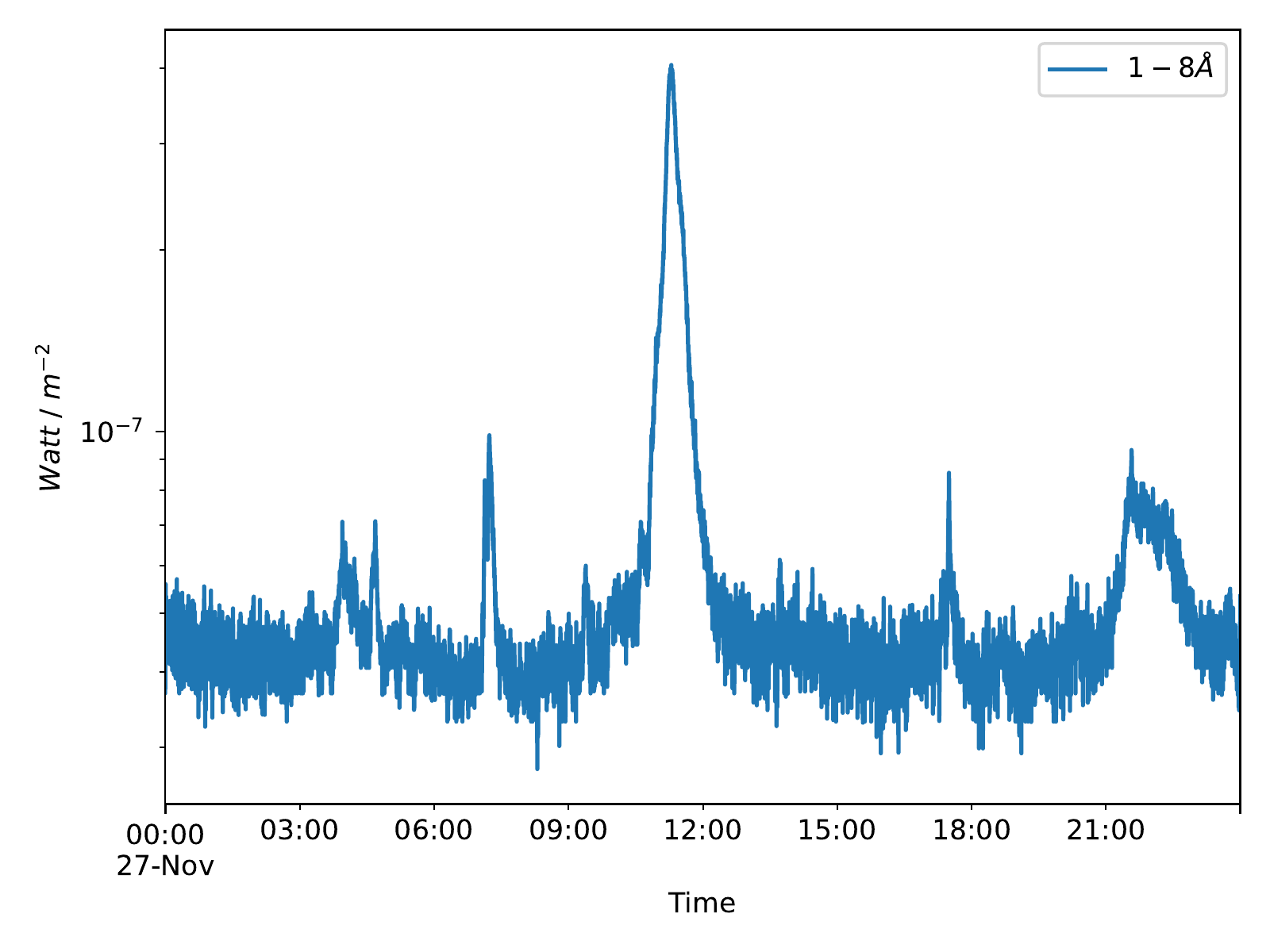}
    \caption{{X-ray light curve from GOES for 27 November, 2017 showing the B2.9 flare and some weaker flares.}}
    \label{fig:goes_lightcurve}
\end{figure}

The data were recorded using the MWA Phase II \citep{tingay2013, wayth2018} on November 27, 2017 under the project ID G0002  from 01:30--03:38 UT.
This day is characterised by a very low level of solar activity\footnote{\url{https://www.solarmonitor.org/?date=20171127}}.
Only one active region (NOAAA 12689) was present on the visible part of the solar disc. {The SWPC catalogue reported a B2.9 class X-ray flare on this day. However, some other smaller flares were also visible in the X-ray light curve from the Geostationary Operational Environmental Satellite (GOES), which is shown in Fig. \ref{fig:goes_lightcurve}. 
No radio flare was reported on this day by the Culgoora spectrograph operating between 18--1800 MHz and Learmonth spectrograph operating between 25--180 MHz.}

\begin{figure}
\centering
    \includegraphics[clip,trim={2cm 0 2.5cm 1cm},scale=0.35]{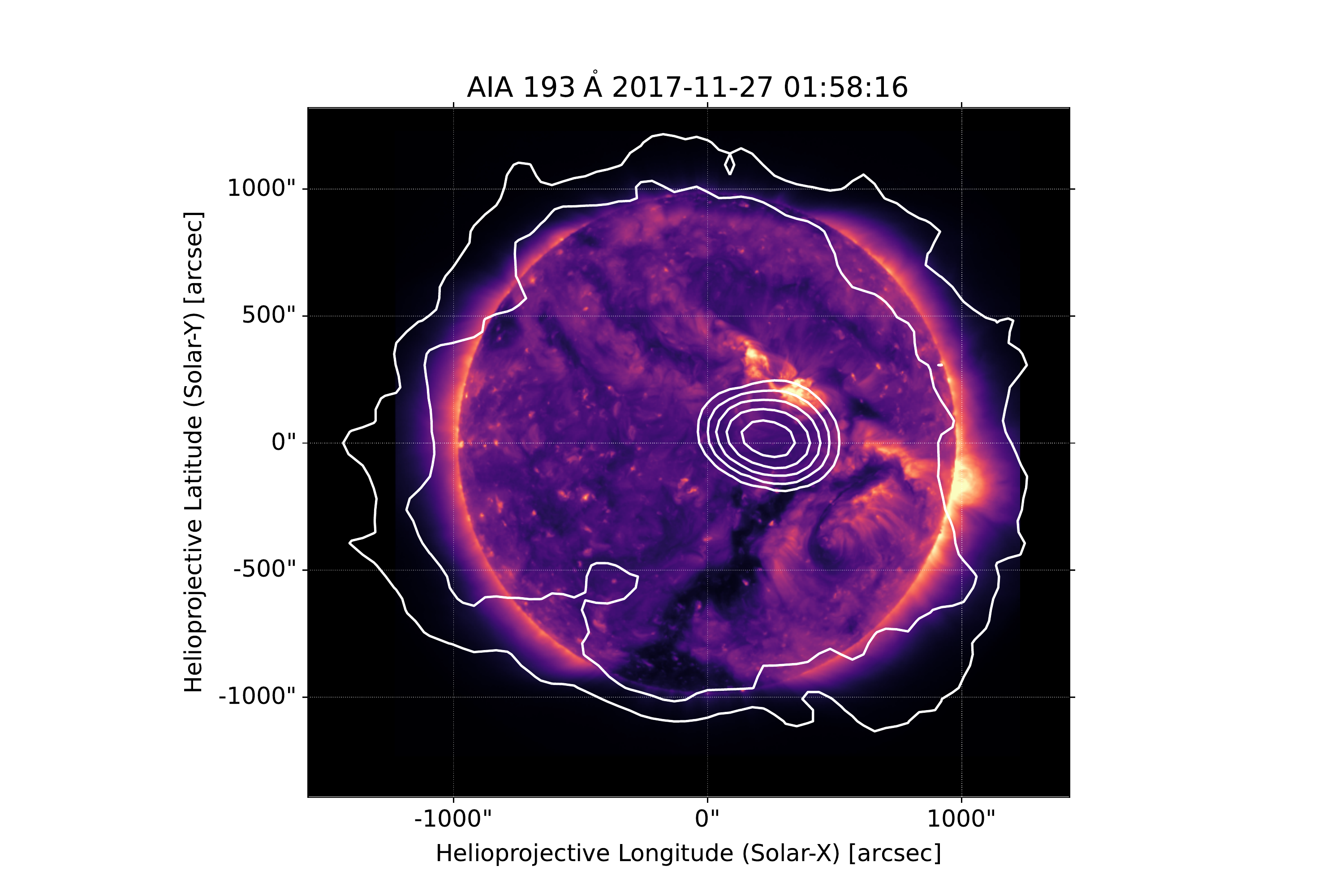}
    \caption{{ An example radio image at 132 MHz is shown in contours which are overlaid on top of an AIA 193\AA $\,$ image taken near the middle of our observing window. The contour levels are at 0.006, 0.01, 0.04, 0.08, 0.16, 0.32, 0.64 times the peak in the radio image. The type I noise storm is very prominent in the radio image and is associated with the lone active region on the solar disc.}}
    \label{fig:solar_image}
\end{figure}

The results presented here come from the time interval 01:30 UT to 02:38 UT.
The observations were done in 12 frequency bands each of 2.56 MHz bandwidth, centred near 80, 89, 98, 108, 120, 132, 145, 161, 179, 196, 217, and 240 MHz.
We have analysed data at four of these frequency bands centered at 108, 120, 132, and 161 MHz.
Imaging was done using the ``Automated Imaging Routine for Compact Arrays of the Radio
Sun" \citep[AIRCARS.][]{mondal2019}, an unsupervised interferometric imaging pipeline developed specifically to meet the challenge of snapshot spectroscopic imaging needs of solar radio science using data from modern arrays. 
Various recent works have established the combination of MWA data and AIRCARS imaging as the state-of-the-art for metrewave solar radio imaging.
Examples include -- 
the detection of QPPs in burst source sizes and orientation
with simultaneous QPPs in intensity for type III and type I solar bursts referred to earlier \citep{mohan2019, mohan2021}; the discovery of 30s QPPs in the type I solar noise storm associated with an active region loop hosting a transient brightening \citep{mohan2019b}; the detection of weak gyrosynchrotron emission from the plasma trapped in the magnetic field of a rather weak Coronal Mass Ejection \citep{mondal2020a};  the discovery of very weak (mSFU level, 1 SFU = 10$^4$Jy) impulsive emissions from the quiet Sun \citep{mondal2020}; and a detailed study of coronal propagation effects under quiet coronal conditions \citep{sharma2020}.
The imaging was done at 0.5 s cadence and 160 kHz frequency resolution, using default parameters. 
The imaging dynamic ranges (DRs) vary significantly across frequency and time. 
The typical DRs at 108, 120, 131 and 160 MHz were 300, 500, 800, 1200. 
An example image is shown in Fig. \ref{fig:solar_image}, where I have overlaid the radio contours on top of an AIA 193\AA $\,$ map. 
A total of about 33,000 such images were analysed for this work.

\section{Results}
\label{sec:inferences}

\begin{figure*}
     \centering
     \includegraphics[trim={2cm 0.5cm 0 0},clip,scale=0.55]{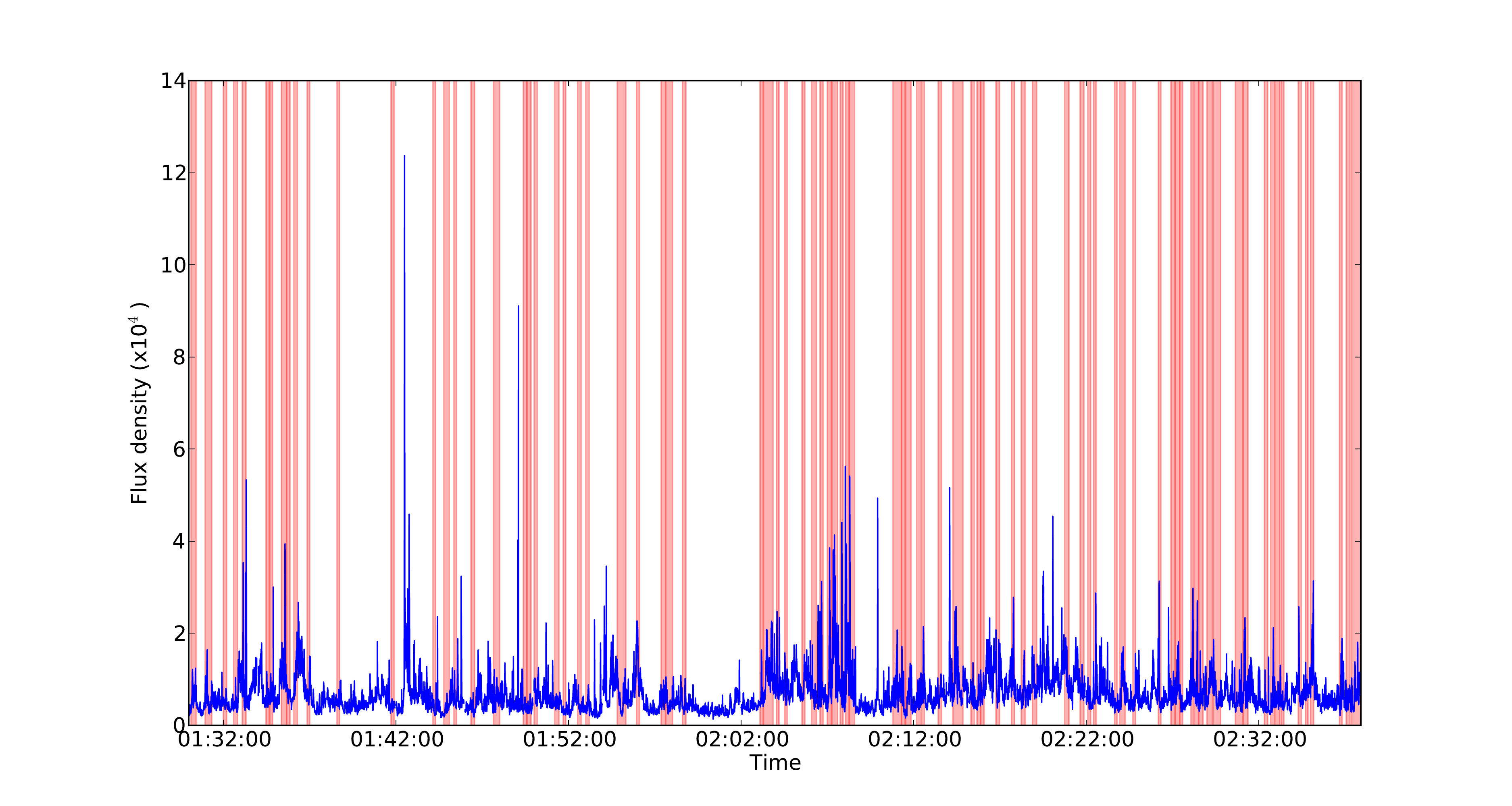}
     \caption{The 132 MHz flux density light curve for the noise storm site. { The light red bands show the times where the anti-correlation between $A_{\nu}$ and $S_{\nu}$ was detected and is discussed later in Sec. \ref{sec:inferences}}
     }
     \label{fig:flux_timeseries}
\end{figure*}

The noise storm seen in top panel of Fig. \ref{fig:solar_image} is associated with NOAA 12689. In the bottom panel, we have also shown the continuum map from the
Helioseismic and Magnetic Imager (HMI) onboard the Solar Dynamics Observatory (SDO). The sunspot associated with the active region is clearly visible.  
The 160 MHz light curve of emission from the site of the noise storm is shown in Fig. \ref{fig:flux_timeseries}.
The coarse flux calibration presented in M20 places the lower envelope of the light curve shown in Fig. \ref{fig:flux_timeseries} at about 0.6 SFU, with the brightest peak close to 01:42:00 corresponding to about 24 SFU.
Even the weakest emission from the noise storm is about 20--30 times brighter than neighboring quiet regions of the solar disc.
The light curve shows some variations over time scales of hundreds of seconds, over which numerous bright impulsive emissions are superposed.
The radio light curves for other frequencies also show similar features.

\begin{figure*}
    \centering
    \includegraphics[trim={2cm 1cm 2cm 1cm},clip,scale=0.35]{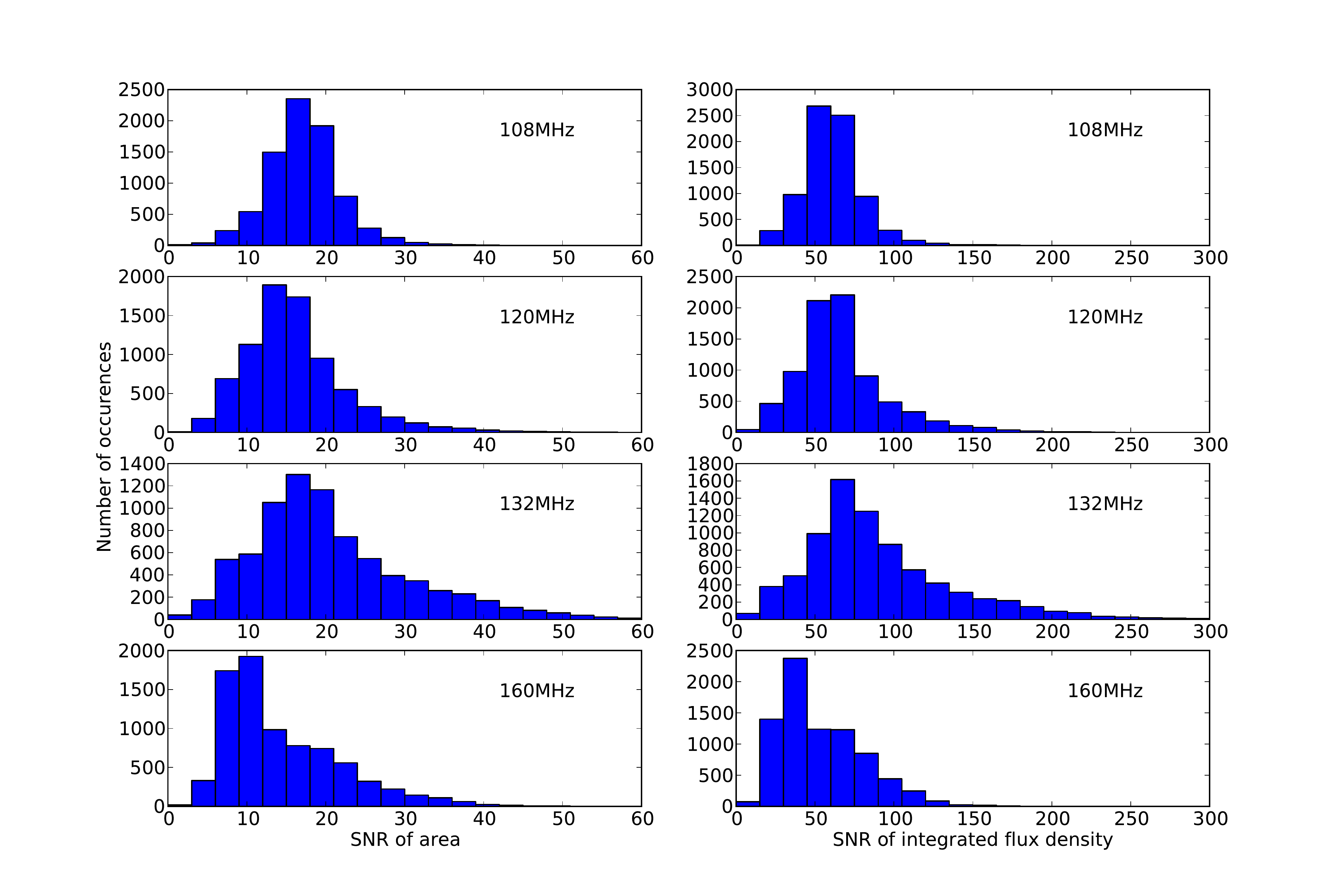}
    \caption{Histograms of signal-to-noise ratio of the estimated deconvolved areas (left column) and integrated flux density (right column).
    }
    \label{fig:snr_histogram}
\end{figure*}

The morphology of the persistent noise storm source is found to be well modelled by a two-dimensional (2D) elliptical Gaussian. 
Using the {\tt imfit} task of Common Astronomy Software Applications \citep[CASA;][]{casa}, 2D Gaussians were fit to noise storm source for all times and frequencies, and the best fit parameters recorded. 
The  restoring beam was deconvolved from the major and minor axis ($\sigma_{major/minor}$) reported by {\tt imfit} to estimate their true values. 
Area of the sources were estimated as $A_{\nu} = \pi\ \sigma_{major}\ \sigma_{minor}$. 
 The volume under the best fit 2D Gaussian is used as the estimate of the integrated flux density of the burst source, $S_{\nu}$.
{\tt imfit} also returns uncertainties on all of the fitted parameters, which were appropriately propagated to calculate the errors in $A_{\nu}$ and $S_{\nu}$. 
Fig. \ref{fig:snr_histogram} shows histograms of signal-to-noise ratio (SNR) of the estimated $A_{\nu}$ and $S_{\nu}$ at times relevant for this work. 
Here SNR refers to the ratio of the fitted value to the associated uncertainty.
Though these are rather weak type I emissions, nonetheless, it is evident from Fig. \ref{fig:snr_histogram} that both these parameters have been estimated with high SNR,  ranging from a minimum of about 10 to a maximum of about 250.

As motivated earlier, our primary objective is to extend earlier work by \citet{mohan2019} and \citet{mohan2021} to the quiescent phase of type I solar bursts by looking for presence of correlations in the evolution of $S_{\nu}$ and $A_{\nu}$ of the best-fit 2D Gaussian source models.
We examined the cross-correlation of the time series of $S_{\nu}$ and $A_{\nu}$ of the burst source for the entire time series, and found no significant correlation.
This implies that these two quantities are not correlated over a major fraction of the time series. 
It does not, however, rule out the possibility that a correlation might well exist over shorter intervals during these $\sim$70 minutes.
To explore this possibility, we searched for such time intervals using the Spearman correlation \citep{spearmanr_ref,spearmanr_ref2}.
The null hypothesis is that the $S_{\nu}$ and $A_{\nu}$ are not correlated. 
{ Spearman correlation coefficient tests for the presence of a {\em monotonic} relationship between two variables,  unlike Pearson's correlation which tests only for the presence of a {\em linear} relationship. We use the function {\it spearmanr} implemented in {\it Scipy} for calculating the Spearman correlation. It returns two parameters: the correlation coefficient, corr, which measures the strength of the correlation and the p-value which quantifies the probability of an uncorrelated system producing datasets which yields Spearman correlation at least equalling that computed from the given dataset.}
We regard an (anti-)correlation to exist over a localized part of the time series if the following conditions are satisfied.

 \begin{enumerate}
     \item The Spearman correlation coefficient is $\leq$ -0.8.
     \item The duration of the time window over which criterion 1 is valid is $>$ 10 s.
     \item The probability of the null hypothesis is $<$ 0.01\%.
 \end{enumerate}
 
 \begin{figure*}
\includegraphics[scale=0.3]{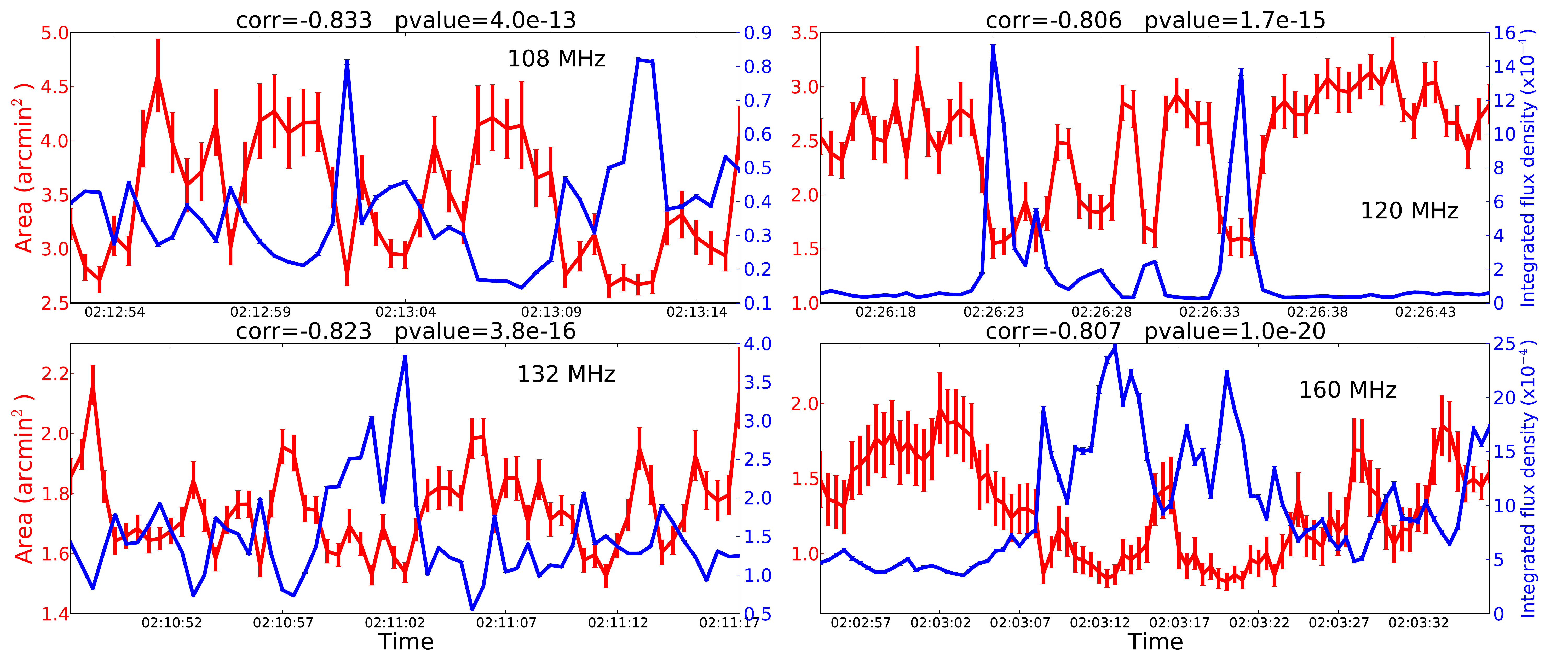}
\caption{The red and blue lines show the $A_{\nu}$ and $S_{\nu}$ of the burst source respectively. The Spearman correlation coefficient (corr) and the probability of the null hypothesis (pvalue) are listed on top of each panel.
}
\label{fig:correlation}
\end{figure*}
 
\begin{figure}
    \centering
    \includegraphics[trim={0.4cm 0.4cm 1cm 1.8cm},clip,scale=0.36]{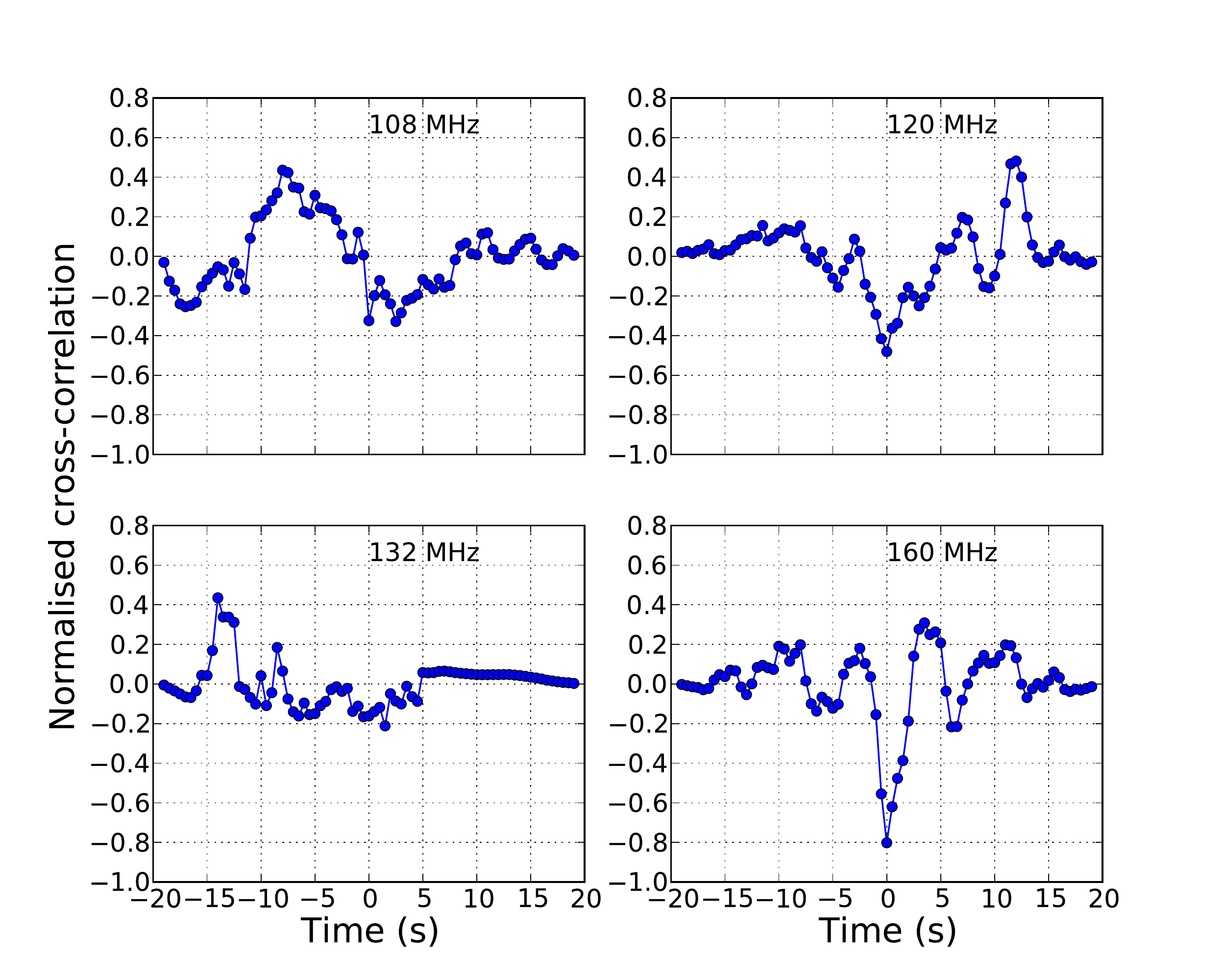}
    \caption{Shows the cross-correlation between $A_{\nu}$ and $S_{\nu}$ at all of the four analysed frequencies at 01:49:01 UT. Unambiguous anti-correlation is observed only at 160 MHz, and a hint for the same is seen at 120 MHz.
    }
    \label{fig:anti_one_freq}
\end{figure}

\begin{figure}
    \centering
    \includegraphics[trim={0.4cm 0.4cm 1cm 1.8cm},clip,scale=0.36]{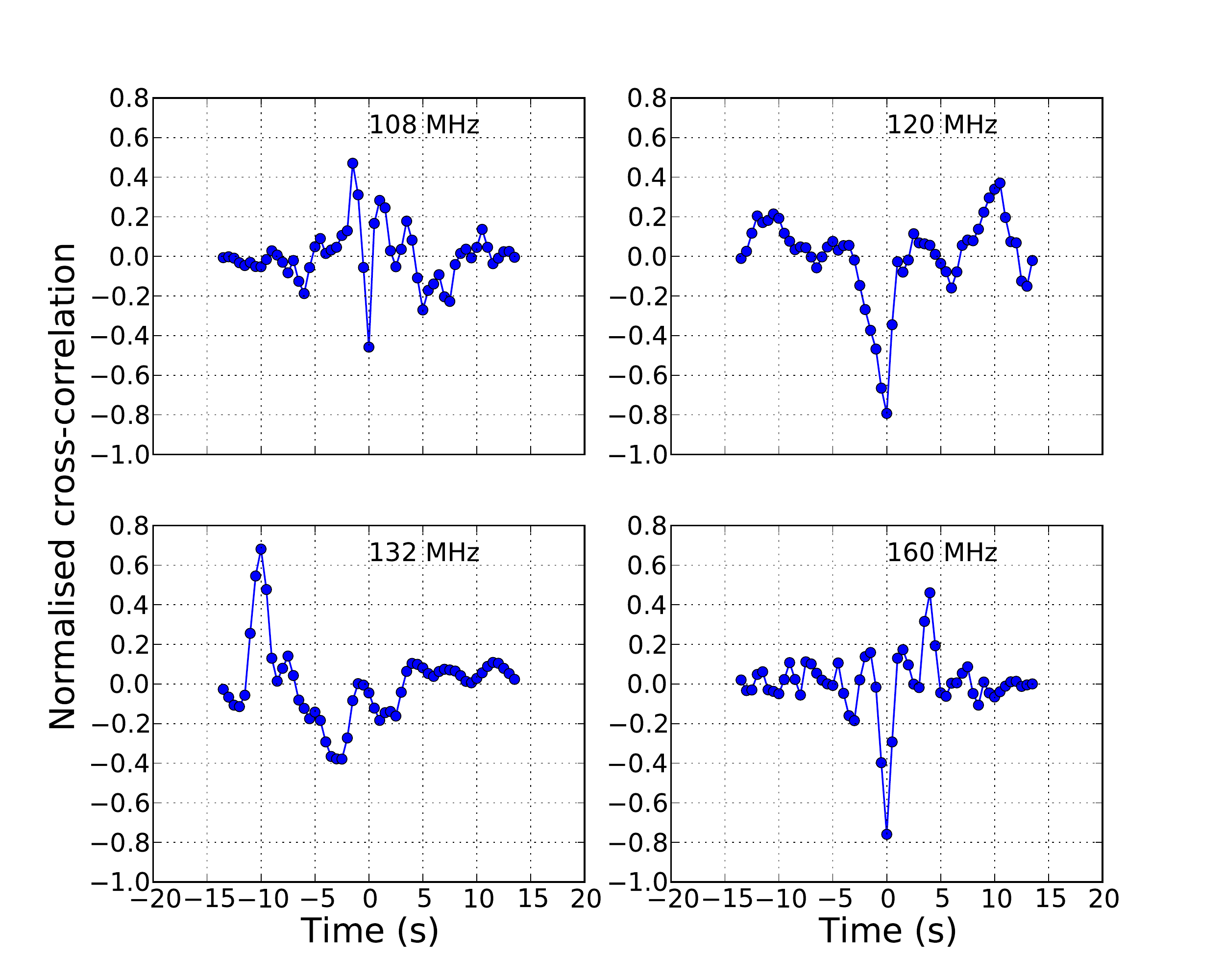}
    \caption{The cross-correlation at 02:12:21 UT in the same format as Fig. \ref{fig:anti_one_freq}.
    Unambiguous anti-correlation is observed at 120 MHz and 160 MHz, and a hint for the same is seen at 108 MHz.
    }
    \label{fig:anti_mult_freq}
\end{figure}

{ To identify instances of presence of anti-correlation, the following procedure was employed.
Starting from some time initial $t_0$, the starting size of the window in which to search for correlation, $\Delta t$, was set to 10 s.
Compute the Spearman coefficient between $S_\nu(t_0:t_0+\Delta t)$ and $A_\nu(t_0:t_0+\Delta t)$, where $(t_0:t_0+\Delta t)$ denotes the time range from $t_0$ to $t_0+\Delta t$.
If the conditions for anti-correlation to exist are satisfied, increase $\Delta t$ by $0.5 s$, continue to do so till the conditions are no longer satisfied and note down the largest $t_0 : t_0 + \Delta t$ for which the conditions were satisfied. 
If not, increase $t_0$ by the larger of $0.5 s$ and $\Delta t$, and carry out the procedure again.
} This analysis for 108, 120, 132 MHz and 160 MHz yielded 27, 71, 84 and 81 instances respectively, where the above conditions were satisfied.
{Figure \ref{fig:flux_timeseries} shows all such instances marked on the 132 MHz flux density time series.}
Figure \ref{fig:correlation} shows four examples, one from each of the frequencies, demonstrating the anti-correlation between $A_{\nu}$ and $S_{\nu}$.  
Though not shown here, in all cases the peak flux density, is found to closely track $S_{\nu}$.
The probability of the null hypothesis being true, pvalue, is also listed and is always found to be $< 10^{-4}$  or 0.01\%.
We also find that the presence of anti-correlation at one frequency does not imply its presence at all of the other frequencies analysed here. 
Figures \ref{fig:anti_one_freq} and \ref{fig:anti_mult_freq} show two examples of times where an unamibguous anti-correlation between $A_{\nu}$ and $S_{\nu}$ is present at some frequencies but not at others.

\begin{figure}
\centering
\includegraphics[scale=0.46]{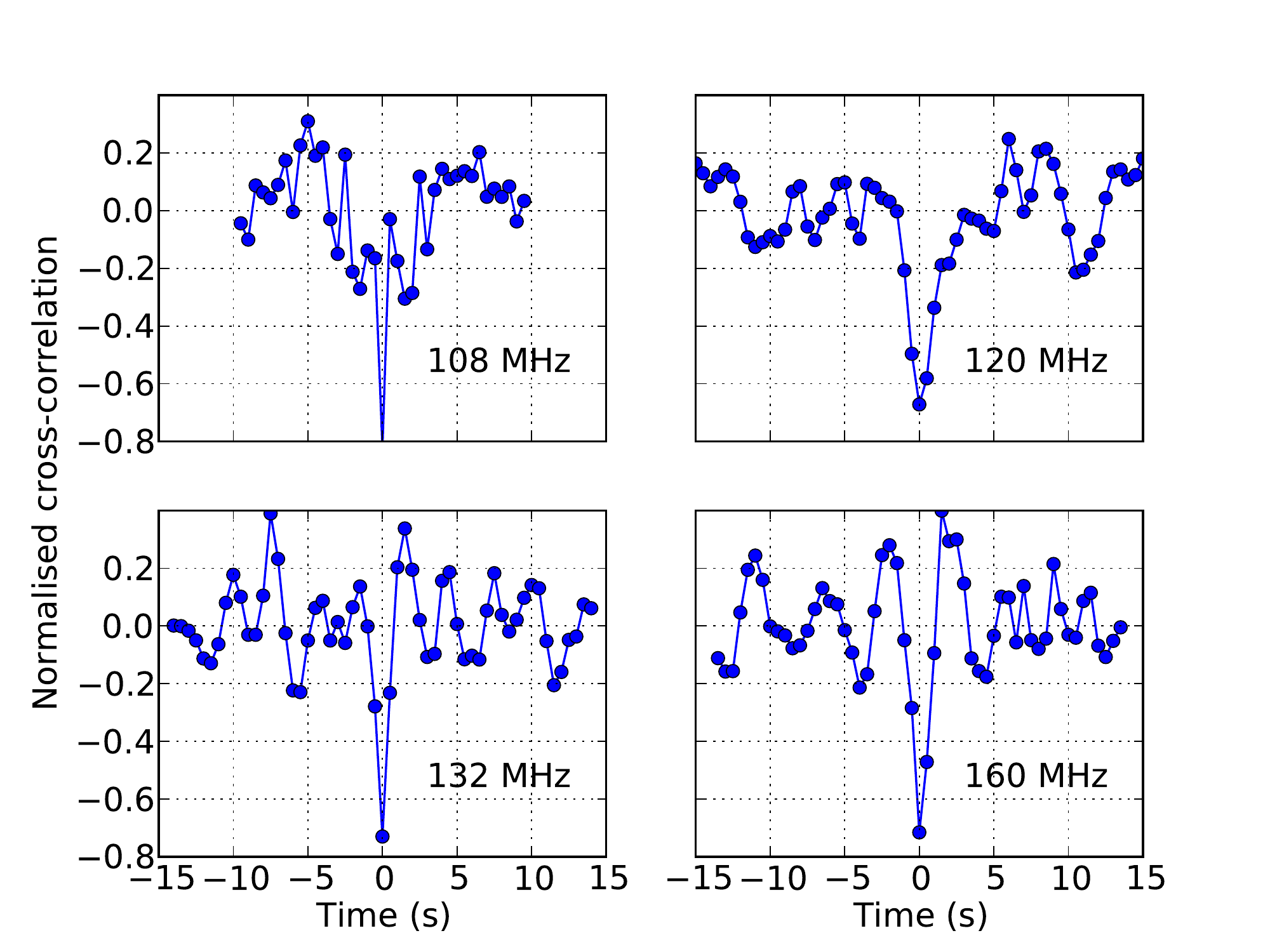}
\caption{The normalized cross-correlation between the $A_{\nu}$ and $S_{\nu}$ at each of the four frequencies for some example time periods. 
Upper left: 108 MHz; Upper right: 120 MHz; Lower left: 132 MHz; Lower right: 160 MHz. 
{The times series corresponding to 108, 120 132 and 160 MHz start at 01:43:49, 02:26:15, 02:26:54 and 01:39:29.5 UT respectively. }}
\label{fig:NCC_day2}
\end{figure}

It is evident from the examples shown in Fig. \ref{fig:correlation}, that many of the time periods show clear signatures of pulsation. 
Figure \ref{fig:NCC_day2} shows an example cross-correlation curve from each of the four frequencies analysed here. {The timeseries and the cross-correlation curves show a hint of periodicities with timescales $\sim 4-10$s, but the present data are insufficient to unambiguously establish their periodic nature.}

\begin{figure}
    \centering
    \includegraphics[scale=0.25]{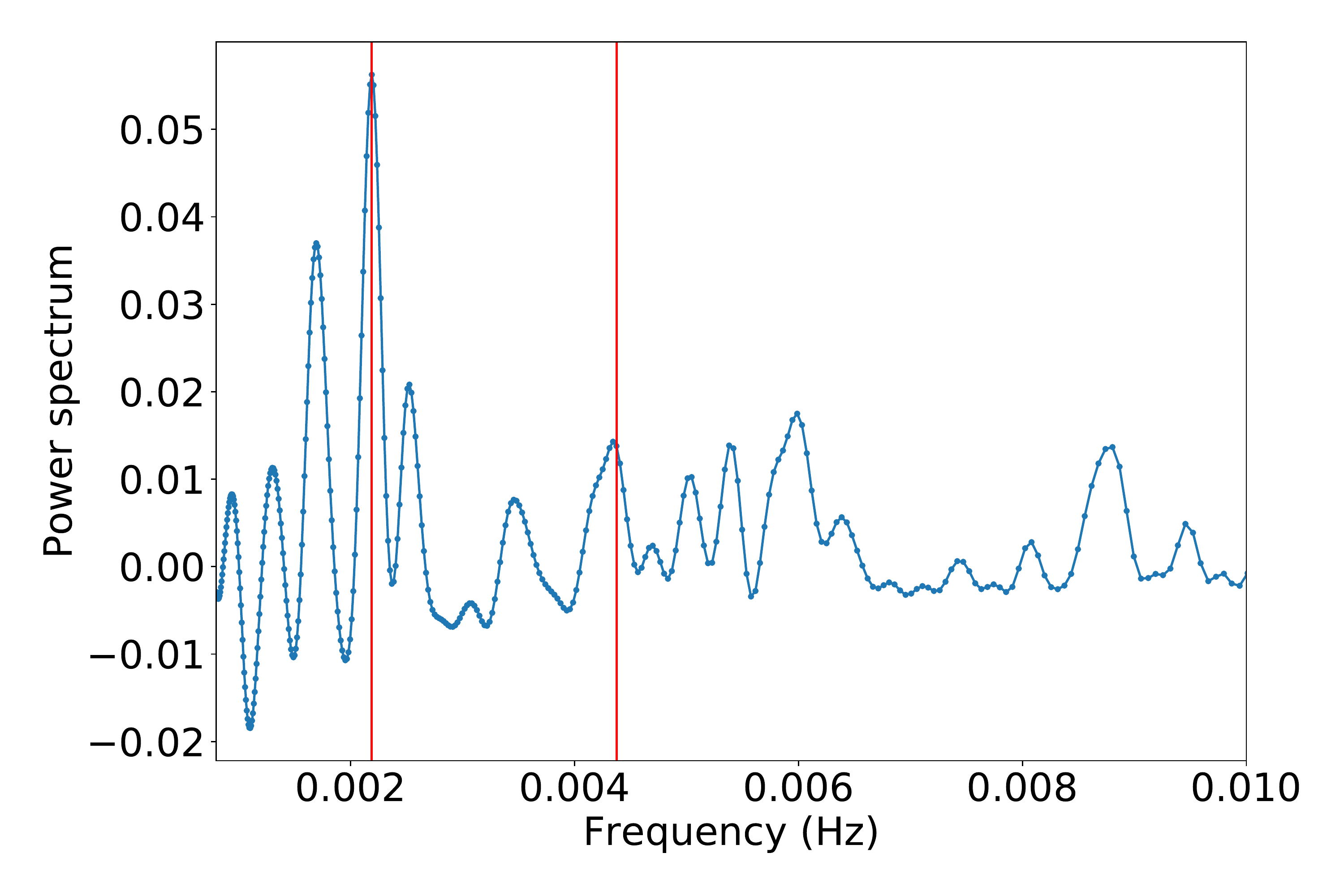}
    \caption{Residual power spectrum at 108 MHz. The red lines are drawn at frequencies corresponding to 0.002188 Hz, which corresponds to timescale of $\sim$ 7 minutes, and its harmonic.}
    \label{fig:108_periodicity}
\end{figure}

To investigate the presence of any long timescale periodicity, we performed a Lomb-Scargle periodogram analysis \citep{vanderplas2018} of the integrated flux density. While Fourier transforming the data and searching for peaks in the power spectrum is routinely used to search for periodicities, the Fourier transform technique cannot take into account measurement errors. Additionally, the fast Fourier transform algorithm routinely used for obtaining the Fourier transform necessarily requires a uniformly sampled grid of measurements which is again not available in general. The Lomb-Scargle periodogram can both handle both non-uniformly sampled data and also take into account measurement uncertainities, making it the right choice for this work.
We only include points where the SNR of the best fit $A_{\nu}$ is $> 10$  and SNR of both $S_\nu$ and peak intensity exceeds 40. The resultant power spectrum has significant red noise component.
To remove the effect of the red noise, we fit a powerlaw to the spectrum and then subtracted it. We searched for peaks in the residual power spectrum. There is a hint of minute scale periodicity for 108 MHz. 
While peaks at similar timescales were present at other frequencies as well, they were not significant given the presence of multiple other peaks of similar strengths.
Figure \ref{fig:108_periodicity} shows the residual power spectrum at 108 MHz. 
To investigate the effect of SNR, the analysis was repeated using data where the SNR of best fit $A_{\nu}$ was $>$ 5 and the SNR of both the integrated flux density and the peak flux was $>$ 10. 
This did not lead any change in the locations nor the strengths of the peaks.

\section{Discussion}

In this paper, we study the time variation of the integrated flux density, $S_{\nu}$, and the area, $A_{\nu}$, of a very weak noise storm over 70 minutes. 
A 2D Gaussian model is found to describe the morphology of this emission well. 
The parameters of the best fit Gaussian are used to infer the deconvolved properties of the noise storm radio source. 
We find robust evidence for multiple time instances when $A_{\nu}$ and $S_{\nu}$ of the noise storm source are anti-correlated with each other over duration exceeding 10 seconds.
From Figs. \ref{fig:correlation} and \ref{fig:NCC_day2}, and some of the panels of Figs. \ref{fig:anti_one_freq} and \ref{fig:anti_mult_freq}, the presence of an anti-correlation between $A_{\nu}$ and $S_{\nu}$ is clearly evident. 
{ The time series for the data used for Figs. \ref{fig:anti_one_freq}, \ref{fig:anti_mult_freq} and \ref{fig:NCC_day2} is shown in the Appendix.}

As mentioned earlier, such a correlation has recently been established for strong type I emission associated with an ARTB during a microflare \cite{mohan2021}.
The emissions under study here, however, are more than two orders of magnitude fainter and, to the best of our knowledge, the faintest by far among earlier works of similar nature. 
Hence we carefully examine the possibility that the observed correlation might arise from an analysis artefact.

\begin{figure}
 \centering
 \includegraphics[trim={1cm 0 0 1cm},clip,scale=0.35]{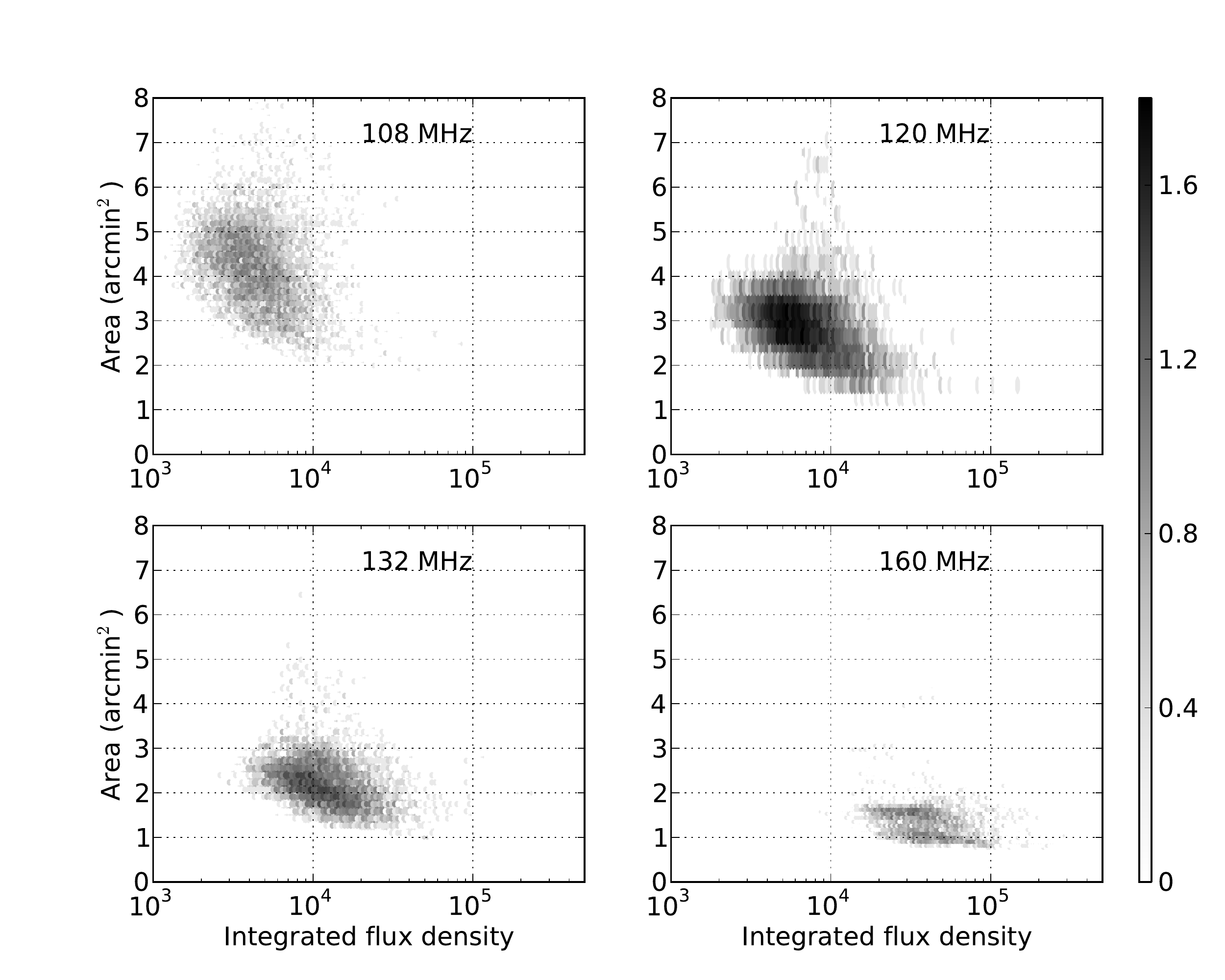}
 \caption{The 2D histogram of $S_{\nu}$ vs $A_{\nu}$ of the best fit Gaussians which satisfy the criteria outlined in the text for each of the four frequencies. The colorbar shows the logarithm of the number of points in each bin.
 }
 \label{fig:for_scattering}
 \end{figure}

\begin{figure}
    \centering
    \includegraphics[trim={1cm 0 0 0},clip,scale=0.4]{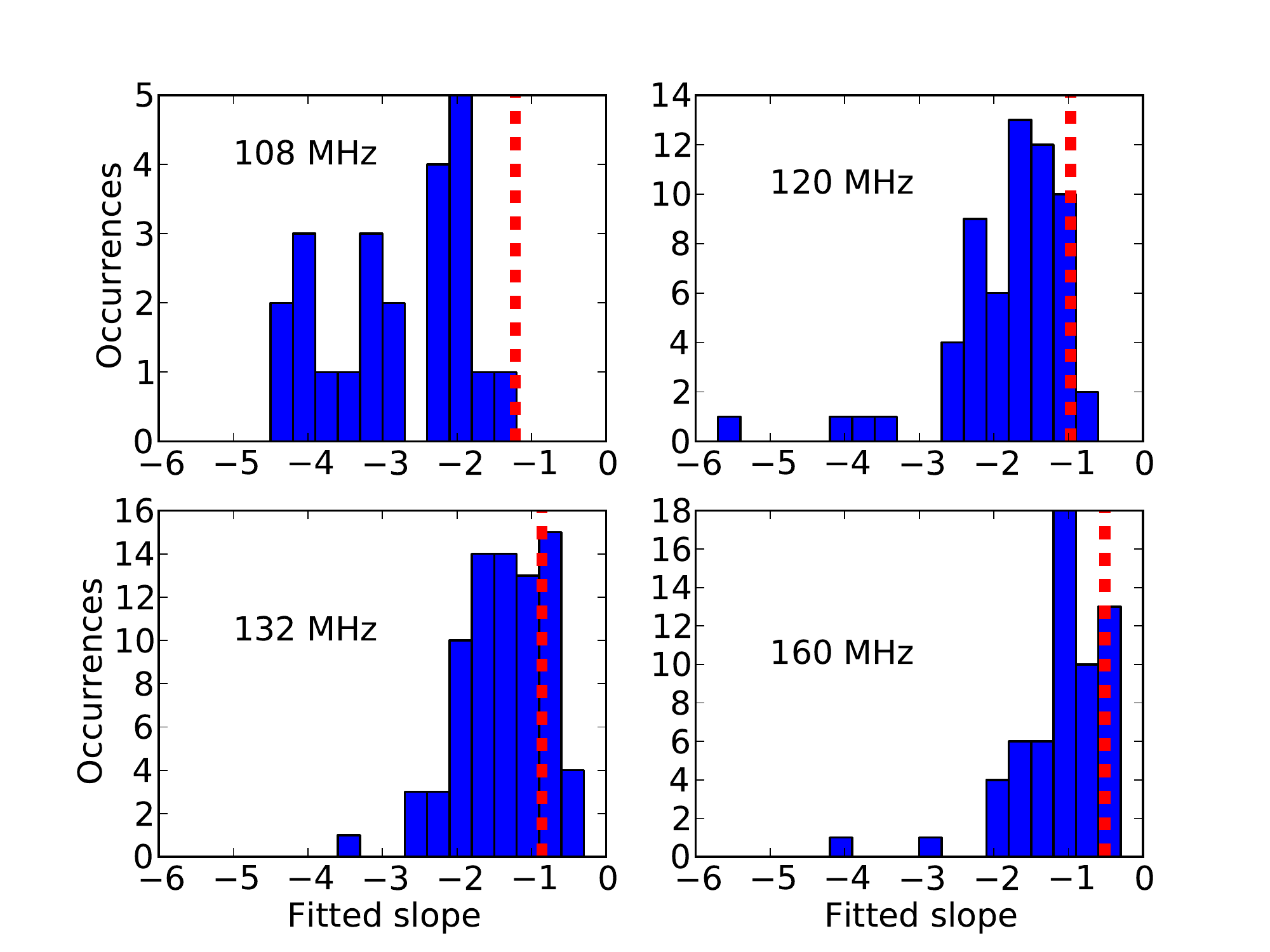}
    \caption{Histogram of the best fit line slopes fitted between $A_\nu$ and $\log_{10} S_\nu$. The red dashed line shows the slope of the best fit line fitted to the points plotted in Fig. \ref{fig:for_scattering}.
    }
    \label{fig:best_fit_slope_val}
\end{figure}

We consider the possibility that such an anti-correlation might arise due to the interplay between the compact source of type I emission, the extended quiet sun emission and the fitting process, leading to larger estimates for best fit sizes (and consequently $A_{\nu}$) when the type I source is weaker, and vice-versa.
Figure \ref{fig:for_scattering} shows a scatter plot of $A_{\nu}$ against $S_{\nu}$ for all of the data for all four frequencies, irrespective of presence of anti-correlation between these two quantities, subject only to the constraints that the SNR of $A_\nu>10$ and SNR of both the fitted peak flux and $S_{\nu}$ is greater than 40.
These thresholds were motivated by the SNRs seen in Fig. \ref{fig:snr_histogram}.
A total of 8154 data points were available for each frequency, of which 7051, 6597, 6904 and 4468 meet this threshold for 108, 120, 132 and 160 MHz respectively.
Except at 160 MHz, the vast majority of the data points meet the SNR threshold.
We can hence expect the distributions seen in Fig. \ref{fig:for_scattering} to be truly representative of these data.
At 160 MHz, an intermittent bright source was present close to the active region under study and its presence lead to poor fits in many instances.
It is evident from Fig. \ref{fig:for_scattering} that while there is a weak anti-correlation between the area and the integrated flux density, there is also a large scatter over this trend.
The Spearman correlation coefficients are -0.33, -0.41, -0.42 and -0.33 for 108, 120, 132 and 160 MHz respectively, much smaller than the threshold of -0.8 used for identifying the incidences of anti-correlation in this paper. 
A straight line fit between $A_\nu$ and $\log_{10}S_\nu$ for the full dataset shown in Fig. \ref{fig:for_scattering} yields best fit slopes of -1.22, -0.97, -0.86 and -0.51 at 108, 120, 132 and 160 MHz respectively.
Figure \ref{fig:best_fit_slope_val} shows histograms of the slopes of the best fit lines between the same quantities, but during the periods when they show a strong anti-correlation. 
It is evident that, at each of the frequencies, the obtained slopes during these times are significantly higher than those for the full dataset. 
We hence conclude that the observed anti-correlation cannot not arise because of an analysis artefact, and truly reflects the dynamics of the noise source.

Recently the presence of such an anti-correlation has been interpreted in terms of regulation of the reconnection process by sausage MHD modes and the subsequent generation of beams of nonthermal electrons \citep{mohan2019,mohan2021}. 
\citet{mohan2019} estimated that the speed with which material must move for the observed second-scale oscillations was about two orders of magnitude faster than the local Alfv\'{e}n speed, and hence ruled out local MHD processes as their cause.
They had studied a group of type III bursts, which are believed arise along open magnetic fields.

Noise storms, on the other hand, are believed to arise from coronal loops.
In view of this we suggest the following physical scenario -- the coronal loops continually get buffeted and perturbed in their turbulent environment.
These perturbations lead to a variety of MHD wave modes being setup in the loops, including sausage modes.
The presence of sausage modes leads to a compression and expansion of the magnetic loop cross-section.
This change in the cross-section leads to a corresponding change in the density of the beam of nonthermal electrons which give rise to the noise storm radio emission.
The higher density of nonthermal electrons gives rise to stronger emission, leading to the observed anti-correlation observed between $A_\nu$ and $S_\nu$.


The low brightness of the noise storm emission and the quiescent state of the active region suggest that these electron beams must be weak.
Weak electron beams are expected to thermalize quickly due to collisions \citep{mohan2019b} and are hence not expected to traverse coronal regions large enough to be able to simultaneously give rise to anti-correlations at the different coronal heights sampled by our observing frequencies. This also implies that the noise storm emission observed at different frequencies (or coronal heights) must arise due to independent locally produced beams of nonthermal electrons. Assuming the Newkirk coronal density model and assuming that the emissions are arising due to the fundamental mode of emission, we estimate that the 108 MHz and 160 MHz emission are arising from a heliocentric height of 1.22 and 1.11 $R_\odot$ respectively. If we assume that the observed emission is coming at the first harmonic of the plasma frequency, then the corresponding heights for observed emissions at 108 and 160 MHz are 1.47 and 1.31 $R_\odot$ respectively.

\citet{mohan2019} favoured a sausage mode regulated reconnection scenario taking place deep down in the corona where the Alfv\'{e}n speed could be sufficiently large to explain the observed oscillations.
This scenario, however, requires the electron beams regulated by this mechanism to rise to coronal heights large enough to give rise to emission at the observed frequencies, rather than being local in origin.
Also if the anti-correlation is observed at some frequency $\nu_1$, it must necessarily be observed at all frequencies $> \nu_1$, which correspond to lower coronal heights.
This is not consistent with some of the observations. 
We also note that while it is indeed a possibility, these data are insufficient to investigate if the sausage modes influence the local reconnection processes responsible for generating the weak nonthermal electron beams.

\section{Conclusion}

In this work, we have studied the dynamics of radio emission associated with a rather weak noise storm source over a $\sim$70 minute period. The flux density of the noise storm source varied between $\sim$0.6--24 SFU, about two orders of magnitude weaker than earlier studies along similar lines. 
We have discovered multiple instances where the integrated flux density of the burst source is anti-correlated with its area. 
We also find that the presence of anti-correlation at one frequency does not necessarily imply its presence at other neighboring frequencies in the same time window. 
We have proposed a self-consistent scenario to explain this observation. 
We suggest that sausage MHD modes are stochastically excited in quiescent coronal loops. 
These sausage modes change the density of nonthermal electrons responsible for the radio emission, thereby producing the observed anti-correlation between the area and the integrated flux density of the noise storm source. 
{ In a future work, we intend to simulate these conditions and compare the observed properties with the results obtained from our simulation.} 
Our observations also suggest that the nonthermal electron beams responsible for the observed radio emission are local in origin.
Our work provides strong evidence that even during very quiescent times, there is discernible magnetic activity in the vicinity of active regions and in coronal loops.
It also suggests that MHD oscillations in coronal magnetic loops and strands are likely quite ubiquitous.
The radio emission from the weak electron beams propagating through these loops and strands serves to `light' them up, allowing us to detect them.

\begin{figure}
    \centering
    \includegraphics[scale=0.35]{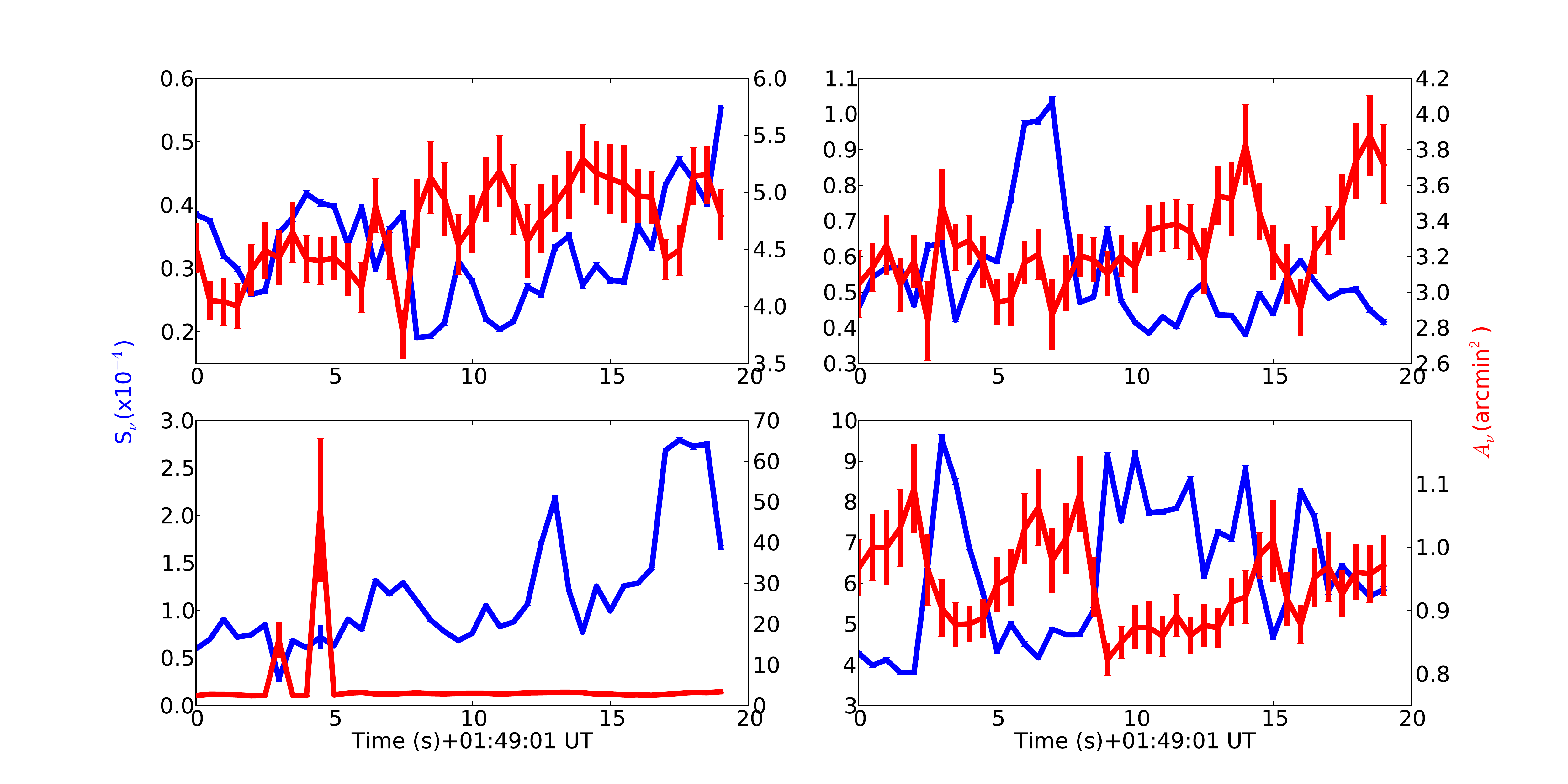}
    \caption{Time series for the data corresponding to Fig.\ref{fig:anti_one_freq}. The blue and red lines indicate $A_\nu$ and $S_\nu$ respectively. Left upper panel:108 MHz, Right upper panel:120 MHz, Left lower panel: 132 MHz, Right lower panel: 160 MHz.}
    \label{fig:fig6-timeseries}
\end{figure}

\begin{figure}
    \centering
    \includegraphics[scale=0.35]{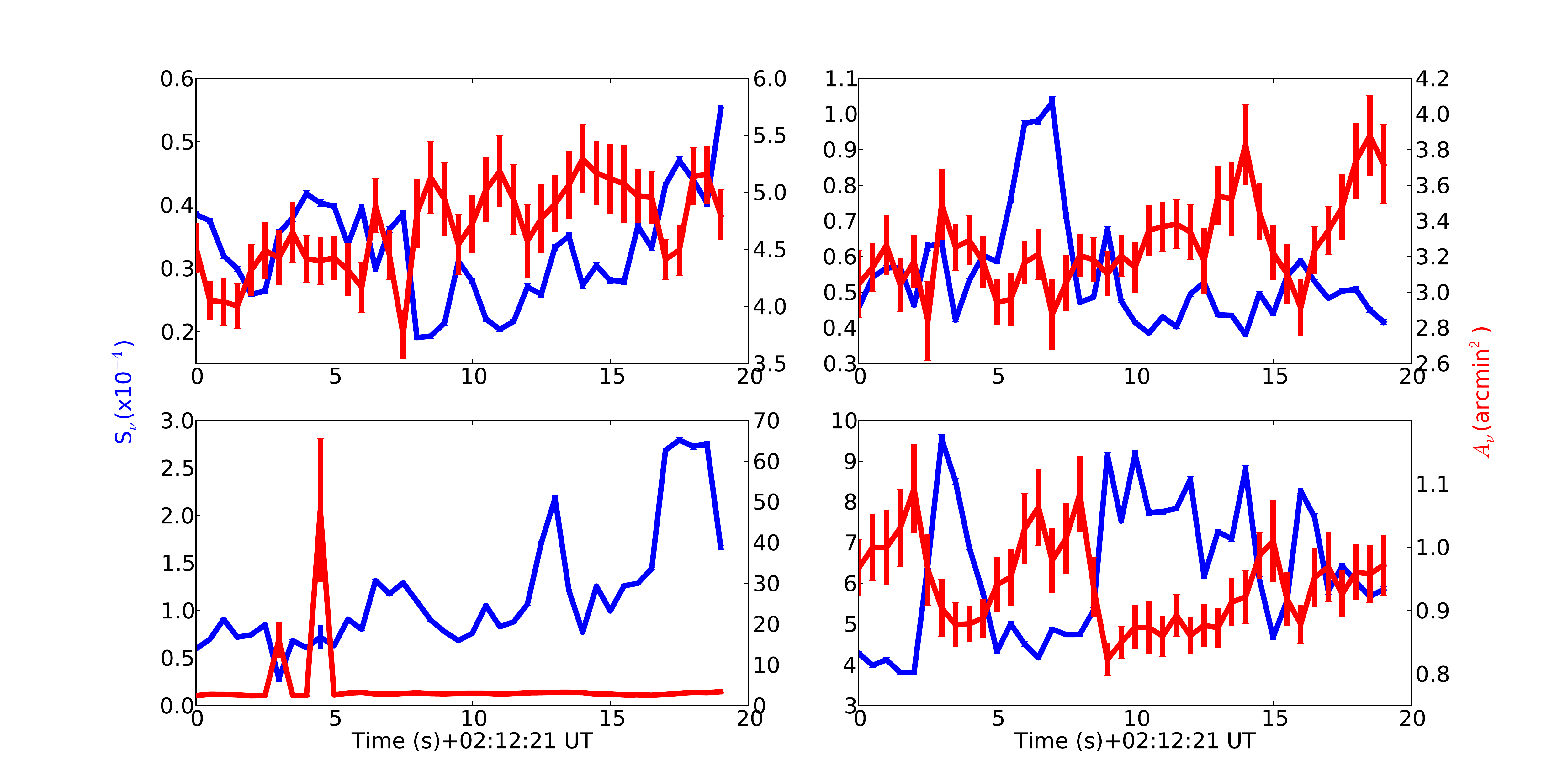}
    \caption{Time series for the data corresponding to Fig.\ref{fig:anti_mult_freq} in the same format as Fig. \ref{fig:fig6-timeseries}. 
    }
    \label{fig:fig7-timeseries}
\end{figure}

\begin{figure}
    \centering
    \includegraphics[scale=0.35]{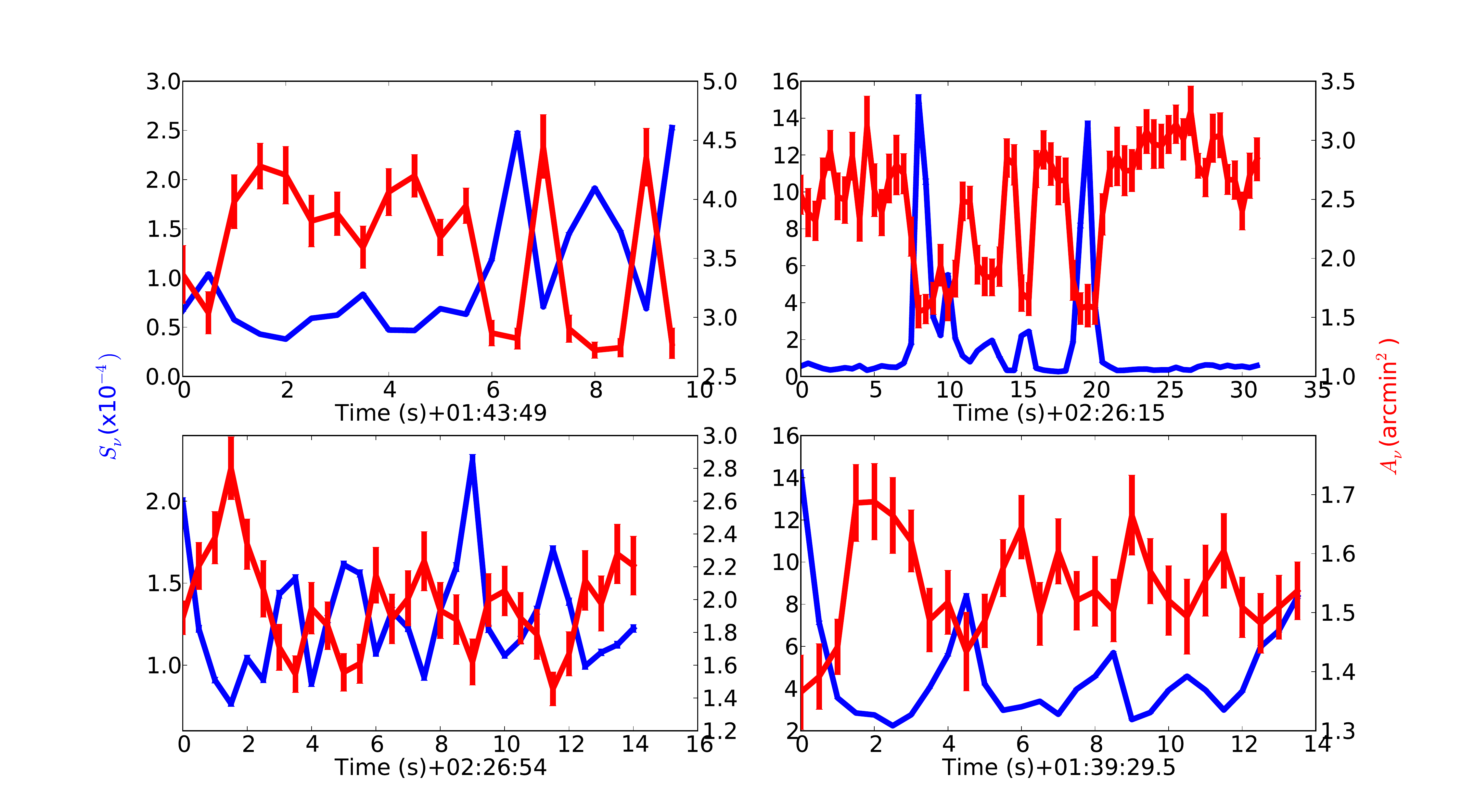}
    \caption{Time series for the data corresponding to Fig.\ref{fig:NCC_day2} in the same format as Fig. \ref{fig:fig6-timeseries}. 
    }
    \label{fig:fig8-timeseries}
\end{figure}

\appendix
{ Figures \ref{fig:fig6-timeseries}, \ref{fig:fig7-timeseries} and \ref{fig:fig8-timeseries} show the time series corresponding the data presented in Figs. \ref{fig:anti_one_freq}, \ref{fig:anti_mult_freq} and \ref{fig:NCC_day2} respectively.
}

\begin{acknowledgements}
This scientific work makes use of the Murchison Radio-astronomy Observatory (MRO), operated by the Commonwealth Scientific and Industrial Research Organisation (CSIRO).
We acknowledge the Wajarri Yamatji people as the traditional owners of the Observatory site. 
Support for the operation of the MWA is provided by the Australian Government's National Collaborative Research Infrastructure Strategy (NCRIS), under a contract to Curtin University administered by Astronomy Australia Limited. We acknowledge the Pawsey Supercomputing Centre, which is supported by the Western Australian and Australian Governments.
We acknowledge support of the Department of Atomic Energy, Government of India, under the project no. 12-R\&D-TFR-5.02-0700.
The SDO is a National Aeronautics and Space Administration (NASA) spacecraft, and we acknowledge the AIA science team for providing open access to data and software. 
This research has also made use of NASA's Astrophysics Data System (ADS). 
We thank the developers of Python 2.7\footnote{See
https://docs.python.org/2/index.html.} and the various associated packages, especially Matplotlib\footnote{See http://matplotlib.org/.}, Astropy,\footnote{See http://docs.astropy.org/en/stable/.} and NumPy\footnote{See https://docs.scipy.org/doc/.}. 
\end{acknowledgements}

%
%

\software{CASA (McMullin et al. 2007), astropy (The Astropy Collaboration 2013, 2018), matplotlib (Hunter 2007), Numpy (Van der Walt, Colbert \& Varoquaux 2011; Harris, Millman \& Oliphant 2020)}

\facility{Murchison Widefield Array, SDO (AIA)}
\bibliography{references}   

\end{document}